# A natural ionic liquid: low molecular mass compounds of aggregate glue droplets on spider orb webs

Yue Zhao [1,2,*], Takao Fuji [1], Masato Morita [3], Tetsuo Sakamoto [3]

[1] Laser Science Laboratory, Toyota Technological Institute, 2–12–1 Hisakata, Tempaku-ku, Nagoya 468–8511, Japan.

[2] Collaborative Open Research Center, Kogakuin University, 2665–1 Nakano, Hachioji, Tokyo, 192–0015 Japan.

[3] Department of Applied Physics, School of Advanced Engineering, Kogakuin University, 2665–1 Nakano, Hachioji, Tokyo, 192–0015 Japan.

ABSTRACT

The aggregate glue of spider orb web is an excellent natural adhesive. Orb-weaver spiders use micron-scale aggregate glue droplets to retain prey in the capture spiral silks of their orb web. In aggregate glue droplets, highly glycosylated and phosphorylated proteins dissolve in low molecular mass compounds. The aggregate glue droplets show a heterogeneous structural distribution after attaching to the substrate. Although components of the aggregate glue droplets have been well analyzed and determined in past studies, visualization of the spatial distribution of their chemical components before and after their attachment is the key to exploring their adhesion mechanisms. Here, we investigated the distribution of low molecular mass compounds and glycoproteins in aggregate glue droplets using the *in situ* measurement methods and visualized the role of specific low molecular mass compounds in promoting glycoprotein modification in the aggregate glue. The results of the analysis suggest that the constituents of aggregate glue droplets include at least one ionic liquid: hydrated choline dihydrogen phosphate, while the modification of glycoproteins in aggregate glue depends on the concentration of this ionic liquid. This natural ionic liquid does not affect the fluorescence activity of fluorescent proteins, indicating that proteins of aggregate glue droplets can be dissolved well and maintain the stability of their higher-order structures in that ionic liquid. As a natural ionic liquid, aggregate glue droplets from the spider orb webs may be an excellent ionic liquid material model.

**Introduction**

Orb-weaver spiders capture their prey by weaving orb webs. Orb-weaver spiders have 7 secretory glands with their own functions. The high-strength silk thread secreted by the major ampullate gland is used as the radial structure skeleton of the orb web. The highly elastic capture line from the flagelliform gland and viscous substances from the aggregate glands are secreted together to form capture spiral lines. Initially, the viscous substances secreted from aggregate glands are uniformly coated on the capture lines, and then due to Plateau-Rayleigh instability [1–3], equidistant micron-scale spherical structures are formed in less than tens of seconds. This structure looks like a beads-on-a-string, and is called aggregate glue droplets. The main components of the aggregate glue droplets are two glycoproteins (ASG-1 and ASG-2) [4–9], low molecular mass compounds (*γ*-aminobutyramide, *N*-acetyltaurine, isethionic acid, choline, betaine, 2-pyrrolidone, putrescine, *N*-acetylputrescine, *ß*-alaninamide and glycine) [10–16], inorganic salts (potassium nitrate and potassium dihydrogen phosphate) [12,14,15], and water [12,13]. The inorganics and low molecular mass compounds dissolved in aggregate glue droplets account for about 25%–40% of the mass of the entire orb web [10,16]. *Araneoid* spiders' ability to synthesize different organic low molecular mass compounds varies [17]. Therefore the spider needs to ingest the necessary organic low molecular mass compounds by recycling previous orb webs [17].

When unfortunate prey hits the orb web, the aggregate glue droplets spread along the object surface within microseconds to seconds after contact with the object, and the glycoproteins in aggregate glue droplets rapidly convert from a viscous to a hardened protein. The microscope images in Fig. 1 show the appearance of the aggregate glue droplets before and after adhesion to the substrate. The aggregate glue droplets show a heterogeneous structural distribution that looks like a "fried egg" after attaching to the substrate (see Fig. 1d). A combination of Raman spectroscopy and light microscopy observations revealed that in the "fried egg" structure of aggregate glue droplets attached to a substrate, the material in the central area is a protein [18]. The results of solid-state nuclear magnetic resonance analysis indicate that the low molecular mass compounds and glycoproteins in aggregate glue droplets play key roles in their adhesive properties [19], that is, the loss of the low molecular mass compounds leads to protein hardening [19,20]. Therefore, a suspended aggregate glue droplet means that the protein is in a dissolved state, while an attached aggregate glue droplet means that the protein is precipitated. In the "fried egg" structure of the attached aggregate glue droplet, low molecular mass compounds and proteins should be separated and complementary in distribution. Choline in low molecular mass compounds is a common ionic liquid cation, and dihydrogen phosphate is a common ionic liquid anion. This suggests that the low molecular mass compounds may be ionic liquids. Here, we assume that the low molecular mass compounds of aggregate glue droplets are ionic liquids that dissolve proteins. If the assumption holds, the choline cation in low molecular mass compounds as the counterion of dihydrogen phosphate should exhibit the same distribution as $Na^+$, $K^+$, and dihydrogen phosphate. In this study, we characterized the inorganic and organic components of aggregate glue droplets at atomic and molecular levels using in situ measurement methods of the time-of-flight secondary ion mass spectrometry (TOF-SIMS), mid-infrared hyperspectral imaging, fluorescence microscopy, second harmonic generation microscopy, and scanning electron microscope energy dispersive X-ray spectrometry. The composition distribution observed by the mapping images of mass spectrometry and infrared absorption spectroscopy confirmed the above hypothesis and provided experimental evidence for the attachment mechanism of aggregate glue droplets.

**Results**

**1. Fluorescence of aggregate glue droplets.**

Figures 1a and 1d are the optical microscopic images of suspended and attached aggregate glue droplets with white light illumination respectively. Figures 1b and 1e are the fluorescence microscopy images of suspended and attached aggregate glue droplets excited by an ultraviolet light source respectively and observed by a CCD camera with a long pass filter with a cut-on wavelength at 420 nm. Figures 1c and 1f are the optical spectra of fluorescence of suspended and attached aggregate glue droplets by a fiber spectrometer, respectively. In Fig. 1d, the opaque granules at the core of each droplet are apparent. The opaque granules are hardened glycoprotein which acts as an anchor to the silk [21]. The optically transparent region around the granule is considered that is a glycoprotein glue [21]. On the other hand, in the fluorescence microscopy image of attached aggregate glue droplets of Fig. 1e, the fluorescence signals were observed from the granules and the glycoprotein glue region. The fluorescence signals from the granules are stronger than that from the glycoprotein glue region. Interestingly, in Fig. 1e, the outer layer of the glycoprotein glue region shows a non-luminous structure. It suggests that this area has no fluorescence activity. While in Fig. 1d, the area corresponding to this outer layer is optically transparent in the direction perpendicular to the substrate and cannot be optically distinguished from the glass substrate. The results of subsequent mass spectrometry mapping confirmed that the non-luminous area in Fig. 1e is the exudate of low molecular mass compounds and inorganic salts. Therefore, the origin of the fluorescent signal in Fig. 1e was determined to be the glycoprotein, namely, the glycoprotein of aggregate glue droplet is a fluorescent protein.

In the suspended aggregate glue droplets, fluorescence is emitted from the whole of the aggregate glue droplet, namely the distribution of glycoprotein in suspended aggregate glue droplets is uniform. Moreover, the fluorescence spectra of suspended and attached aggregate glue droplets are consistent (see Figs. 1c and 1f). In Fig. 1c, two spectral components and their sums are shown. The spectral shapes were well reproduced by a Voigt-fitting procedure with two peak components. The center wavelength of spectral component 1 is 500 nm and spectral component 2 is 536 nm. Hence, at least two chromophores are possessed in the aggregate glue droplet. The spectra of spectral components 1 and 2 are close to the spectra of green fluorescent protein (GFP) and yellow fluorescent protein (YFP), respectively. The chromophore in aggregate glue droplet seems to the GFP-like proteins or the same protein family as GFP. In order to clarify the details of the non-luminous area of the attached aggregate glue droplets, it is necessary to analyze its composition.

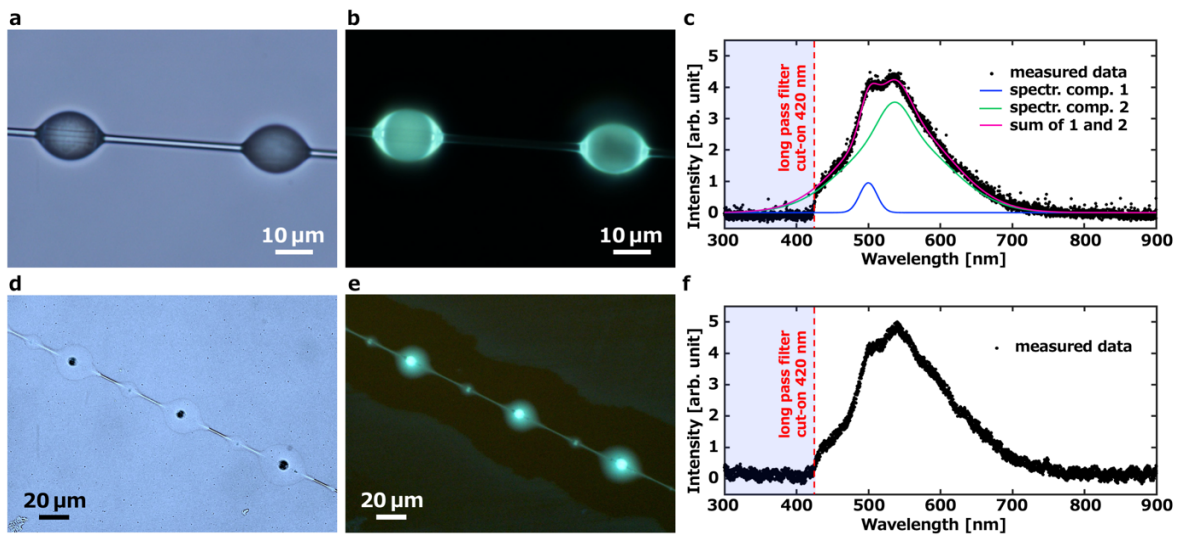

**Figure 1 | Fluorescence observations of aggregate glue droplets. a,** Optical microscopy image of suspended aggregate glue droplets by illuminating with white light. **b,** Fluorescence microscopy image of suspended aggregate glue droplets. **c,** Fluorescence spectrum of suspended aggregate glue droplets. The spectrum was fit with Voigt functions for spectral component 1 (blue line) and spectral component 2 (green line). The magenta line is the sums of the spectral component 1 and 2. **d,** Optical microscopy image of aggregate glue droplets attached to a glass substrate by illuminating with white light. **e,** Fluorescence microscopy image of aggregate glue droplets attached to a glass substrate. **f,** Fluorescence spectrum of aggregate glue droplets attached to a glass substrate.

## 2. Characterization of low molecular mass compounds

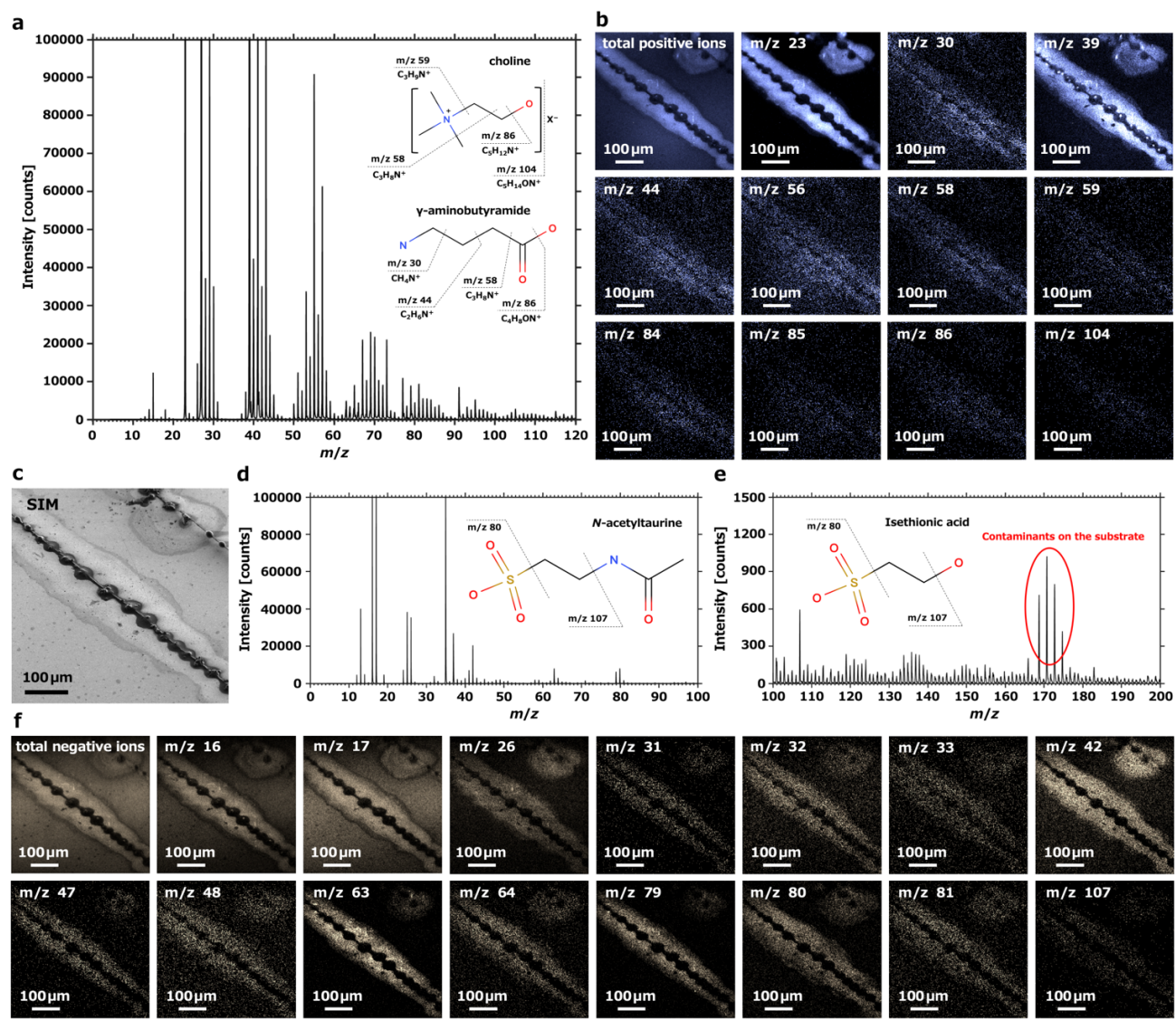

**Figure 2 | Observations of aggregate glue droplets using TOF-SIMS. a,** Positive ions mass spectrum of *m/z* 1 to 120, and **b,** secondary ion mapped images of positive ions. **c,** Scanning ion microscopy (SIM) image of the aggregate glue droplets attached to the substrate. Negative ions mass spectra of **d,** *m/z* 1 to 100 and **e,** *m/z* 100 to 200. **f,** Secondary ion mapped images of negative ions. The mapped color is a pseudo-color.

TOF-SIMS is an excellent surface analysis method. In particular, TOF-SIMS with a focused Ga$^+$ ion beam is effective for analyzing small particles [22]. Hence, here the sample of attached aggregate glue droplets was mapped its element distribution using a home-made TOF-SIMS. We measured positive and negative ions separately with a mass spectrometry range of *m/z* 1 to 300 (see Supporting Information Figs. S1 and S2). First, we observed the scanning ion microscopy image of the sample (see Fig. 2c). In the scanning ion microscopy image, two capture spiral lines were observed, the diagonal one in the field of view was completely attached to the substrate, and the upper right one was partially suspended. The attached aggregate glue droplets took on a "fried egg" structure. From the contrast of scanning ion microscopy image, the "fried egg" structure consists of a darker central area and a spread brighter area.

In all mapping images, we selected images which intensity distribution can coincide with the "fried egg" structure and showed them in Figs. 3d and 3f. All mapping images of positive and negative ions show in Supporting Information Figs. S3 and S4.

In the positive secondary ion mass spectrum and mapped images in Figs. 3a and 3b, *m/z* 23 and *m/z* 39 are assigned to $Na^+$ and $K^+$, respectively. This is consistent with our previous analysis of the inorganic matter of the aggregate glue droplets [20,23]. Other fragment ions are attributed as follows: *m/z* 30 corresponds to $(H_2N–CH_2)^+$, *m/z* 44 to $(H_2N–CH_2–CH_2)^+$, *m/z* 58 to $(H_2N–CH_2–CH_2–CH_2)^+$ and/or $((CH_3)_2N=CH_2)^+$, *m/z* 59 to $((CH_3)_3N)^+$, *m/z* 84 to $(C_4H_6NO)^+$, *m/z* 86 to $(H_2N–CH_2–CH_2–CH_2–C=O)^+$, $((CH_3)_3N–CH=CH_2)^+$ and/or $(C_4H_8NO)^+$, and *m/z* 104 to $((CH_3)_3N–CH_2–CH_2–OH)^+$.

The associated fragment ions at *m/z* 30 $(H_2N–CH_2)^+$, *m/z* 44 $(H_2N–CH_2–CH_2)^+$, *m/z* 58 $(H_2N–CH_2–CH_2–CH_2)^+$ and *m/z* 86 $(H_2N–CH_2–CH_2–CH_2–C=O)^+$, provide significant structural information. The $(H_2N–CH_2)^+$, $(H_2N–CH_2–CH_2)^+$ and $(H_2N–CH_2–CH_2–CH_2)^+$ fragment ions originate from the cleavage of the carbon-backbone, and $(H_2N–CH_2–CH_2–CH_2–C=O)^+$ fragment ions originate from the cleavage of C–O bond of the *γ*-aminobutyramide. The intensity distribution of their mapping images in Fig. 3b corresponds to low molecular mass compounds. Therefore, fragment ions at *m/z* 30, *m/z* 44, *m/z* 58 and *m/z* 86 can be assigned to *γ*-aminobutyramide. Moreover, since the peaks at *m/z* 56 and *m/z* 85 appeared in the mass spectrum of γ-aminobutyramide by electron impact ionization [24], the secondary ion maps of $m/z$ 56 and $m/z$ 85 in Fig. 3b are also assigned to γ-aminobutyramide. In addition, $m/z$ 30 $(H_2N–CH_2)^+$ represents fragment from the cleavage of the C–C bond of glycine, $m/z$ 30 $(H_2N–CH_2)^+$ and $m/z$ 44 $(H_2N–CH_2–CH_2)^+$ represent fragments from the cleavage of the carbon-backbone of ß-alaninamide, $m/z$ 30 $(H_2N–CH_2)^+$, $m/z$ 44 $(H_2N–CH_2–CH_2)^+$ and $m/z$ 58 $(H_2N–CH_2–CH_2–CH_2)^+$ represent fragments from the cleavage of the carbon-backbone of the N-acetylputrescine and putrescine, hence associated fragment ions can also be assigned to glycine, ß-alaninamide, N-acetylputrescine and putrescine. On the other hand, the fragment ions at $m/z$ 58 $((CH_3)_2N=CH_2)^+$, $m/z$ 59 $((CH_3)_3N)^+$, and $m/z$ 86 $((CH_3)_3N–CH=CH_2)^+$ represent the choline residues [25,26]. A peak found at a nominal mass of 104 $m/z$ $((CH_3)_3N–CH_2–CH_2–OH)^+$ represents the molecular fragment of choline [25,26]. Since $m/z$ 58 $((CH_3)_2N=CH_2)^+$ and $m/z$ 59 $((CH_3)_3N)^+$ also represent the betaine residues, associated fragment ions can also be assigned to betaine. From the mapping images in Fig. 3b it is clear that the intensity distributions of $m/z$ 58, $m/z$ 59, $m/z$ 86, and $m/z$ 104 correspond to low molecular mass compounds. The fragment ion at $m/z$ 84 $(C_4H_6NO)^+$ was assigned to the 2-pyrrolidone [27]. The $(C_4H_6NO)^+$ fragment ion originates from the cleavage of the N–H bond of the pyrrolidone ring.

In the secondary ion mass spectrum and mapped images of negative ions in Figs. 3d, 3e and 3f, the typical negative ions at $m/z$ 16 $O^-$, $m/z$ 17 $(OH)^-$ and $m/z$ 32 $O_2^-$ show a consistent distribution with low molecular mass compounds arising mainly from the inorganic and organic salt. The fragment ions at $m/z$ 26 $(CN)^-$ and $m/z$ 42 $(CNO)^-$ assigned to organic nitrogen. On the other hand, although fragment ions $(SO_3)^-$ and $(HPO_3)^-$ have an isobaric interference at $m/z$ 80, the ion mapping at $m/z$ 107 $(SO_3–CH_2–CH_2)^-$ supports that $m/z$ 80 and $m/z$ 107 are from N-acetyltaurine [28] and/or isethionic acid. The negative ion at $m/z$ 33 $(SH)^-$ seems to come from the sulfur in N-acetyltaurine and/or isethionic acid and/or cysteine.

The fragment ions at $m/z$ 31 $P^-$, $m/z$ 47 $(PO)^-$, $m/z$ 48 $(HPO)^-$, $m/z$ 63 $(PO_2)^-$, $m/z$ 64 $(HPO_2)^-$, $m/z$ 79 $(PO_3)^-$, $m/z$ 80 $(HPO_3)^-$ and $m/z$ 81 $(H_2PO_3)^-$ assigned to dihydrogen phosphate. The dihydrogen phosphate is from potassium dihydrogen phosphate [20,23]. Fujita *et al.* synthesized hydrated choline dihydrogen phosphate by adding appropriate amounts of water to choline dihydrogen phosphate [29,30]. Synthesized hydrated choline dihydrogen phosphate as a protein solvent showed a superior affinity with proteins [29–31]. In the synthesized hydrated choline dihydrogen phosphate containing 20 wt% of water, about three water molecules are hydrated into a pair of ions, and no free water molecules exist under this water content [29]. On the other hand, most of the water molecules in the aggregate glue droplets are also bound water [32]. Here, cholinium cations and dihydrogen phosphate anions were detected in the aggregate glue droplets. Therefore, in the aggregate glue droplets, cholinium cations and dihydrogen phosphate anions will spontaneously form hydrated ionic liquids. The water in the aggregate glue droplets will participate in the formation of hydrated choline dihydrogen phosphate, which leads to the low molecular mass compounds have an excellent solubilizing ability for proteins.

Supporting Information Figs. S3 and S4 show all mass spectrum mapping images. In the positive mass spectra, the peak of *m/z* 27 $Al^+$ is from the substrate, while the negative mass spectra peaks of *m/z* 169, *m/z* 171, *m/z* 173, and *m/z* 175 are from the contamination. While each of these low molecular mass compounds has been detected in multiple species, the composition of low molecular mass compounds is different between different species, even between congeneric species [10]. However, in this study, significant differences in the composition of low molecular mass compounds were not found in the sample of *Neoscona nautica*.

Finally, we should note that since the ionization efficiency of low molecular mass compounds is lower than that of inorganic compounds, the beam current (DC) in this study is 5 times higher than that of our previous analysis [23] of inorganic compounds of aggregate glue droplets. The low molecular mass compounds with relatively low ionization efficiency require high beam current, but also pay attention to the damage caused by excessive ion dose. In this study, the beam current (DC) was set to 1.8 nA. In TOF-

SIMS analyses, the dose in pulsed mode was $5.53 \times 10^{12}$ ion/cm$^2$. When the beam current (DC) exceeds 2.4 nA (dose: $7.37 \times 10^{12}$ ion/cm$^2$), both suspended and attached aggregate glue droplets form topographic nanopatterns (see Supporting Information Figs. S5). The topographic nanopatterns depends on the surface material, species of primary ion beam, accelerating voltage of ion beam and beam incident angle [33]. Although this phenomenon can be applied in many fields [34], it needs to be avoided in this study.

## 3. Molecular identification of ionic liquid

To characterize the composition of aggregate glue droplets at the molecular level, we measured the mid-infrared (MIR) transmittance spectra of suspended and attached aggregate glue droplets with a homemade mid-infrared hyperspectral chemical imaging system [35], and mapped and components by spectral angle mapper analysis and linear discriminant analysis.

Hyperspectral imaging and mass spectral mapping as an *in situ* measurement method does not require sample pretreatment and therefore provides more information than bulk analysis. Figs. 3a and 3c are the images of suspended and attached aggregate glue droplets which illuminated by visible light. Figures 3b and 3d are the components mapped images. Figure 3e shows MIR transmittance spectra of the different components which are extracted from the hyperspectral chemical images.

In the mapped image of the suspended aggregate glue droplet of Fig. 3b, a local component distribution was not observed. The MIR transmittance spectrum of the suspended aggregate glue droplet is shown in Fig. 3e upper row. The MIR spectrum of the suspended aggregate glue droplet exhibits a relatively low transmittance due to its large light path length (diameter). Nonetheless, the amide I band (1600−1700 cm$^{-1}$) attributed to the C=O vibration of the main chain of the peptide and the amide II band (1500−1560 cm$^{-1}$) due to the C−N stretching and N−H bending vibrations were observed. The mapping image of Fig. 3b suggests that the distribution of proteins is uniform in the suspended aggregate glue droplets.

In the case of the attached aggregate glue droplet, the attachment leads to the low molecular mass compounds spread along the substrate, and the proteins harden. In mapped image of attached aggregate glue droplet of Fig. 3d, a heterogeneous structural distribution, *i.e.* "fried egg" structure is obvious. Here, two different components were classified and mapped by spectral angle mapper analysis. The middle and bottom row of Fig. 3e show MIR transmittance spectra of two different components which are extracted from the hyperspectral image. The characteristic markers for protein of amide I and II were observed in the central area of the "fried egg" structure. In the MIR transmittance spectrum of spread out area of the "fried egg" structure, characteristic markers of choline dihydrogen phosphate were observed. Choline dihydrogen phosphate is identified using attenuated total reflection Fourier-transform infrared by the vibrational mode 925 cm$^{-1}$ ($\nu_s$[P(OH)$_2$]), 1057 cm$^{-1}$ ($\nu_s$[(PO)$_2$]), 1122 cm$^{-1}$ ($\nu_{as}$[(PO)$_2$]) and 1233 cm$^{-1}$ ($\delta$[OH]) of the dihydrogen phosphate anion [36] and vibrational mode 1478 cm$^{-1}$ (N-C bond) of choline [37]. Since anomalous dispersion in the refractive index of the sample and prism can cause a red shift in the absorption spectrum, the absorption spectrum of choline dihydrogen phosphate measured by the attenuated total reflection method was corrected. As shown in Supporting Information Fig. S6, the dihydrogen phosphate anion vibrational modes were corrected to 957 cm$^{-1}$ ($\nu_s$[P(OH)$_2$]), 1086 cm$^{-1}$ ($\nu_s$[(PO)$_2$]), 1140 cm$^{-1}$ ($\nu_{as}$[(PO)$_2$]) and 1248 cm$^{-1}$ ($\delta$[OH]), and the choline vibrational mode to 1488 cm$^{-1}$ (N-C bond). Choline dihydrogen phosphate is an ionic liquid. Protein can retain its secondary structure after being dissolved by choline dihydrogen phosphate ionic liquid [29].

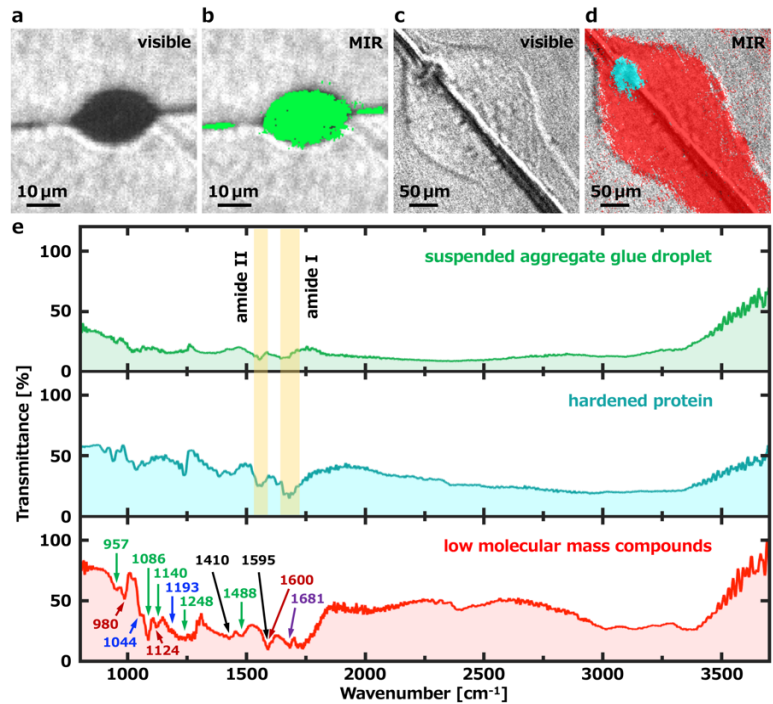

**Figure 3 | MIR hyperspectral imaging and mapping of the aggregate glue droplet.** Suspended aggregate glue droplet images of **a,** visible light illumination and **b,** mapped hyperspectral image. Attached aggregate glue droplet images of **c,** visible light illumination and **d,** mapped hyperspectral image. **e,** Infrared transmittance spectra of the suspended aggregate glue droplet, central and spread out area of attached aggregate glue droplet by extracted hyperspectral image. The mapped color is a pseudo-color.

Since low molecular mass compounds are a complex mixture, its spectrum is the sum of several kinds of substances. Except for choline dihydrogen phosphate, according to known constituents [10–16], the peaks at 1410 cm$^{-1}$ ($v_s$[COO$^-$]) and 1595 cm$^{-1}$ ($v_{as}$[COO$^-$]) are attributed to the carboxylate anion of γ-aminobutyramide, betaine and glycine [38], and the peak at 1681 cm$^{-1}$ is attributed to the hydrophilic carbonyl group in the pyrrole ring of 2-pyrrolidone [39]. The presence of the sulfonate salt is seen by the O=S=O stretching vibrations at 1044 cm$^{-1}$ ($v_{as}$[O=S=O]) and 1193 cm$^{-1}$ ($v_s$[O=S=O]) [40,41]. The sulfonic acid functionality should be from *N*-acetyltaurine and isethionic acid. The NH$_2$ vibrations [42,43] at 980 cm$^{-1}$ (NH$_2$ twisting), 1124 cm$^{-1}$ (NH$_2$ wagging) and 1600 cm$^{-1}$ (NH$_2$ scissoring) are attributed to amino group of putrescine, *N*-acetylputrescine, *ß*-alaninamide and *γ*-aminobutyramide. The hydrogen phosphate dianions can interact with putrescine to form a hydration system [43]. In aggregate glue droplets, putrescine dihydrogen phosphate may also perform dehydration and rehydration functions similar to choline dihydrogen phosphate.

## 4. Direct measurement of adhesion processes

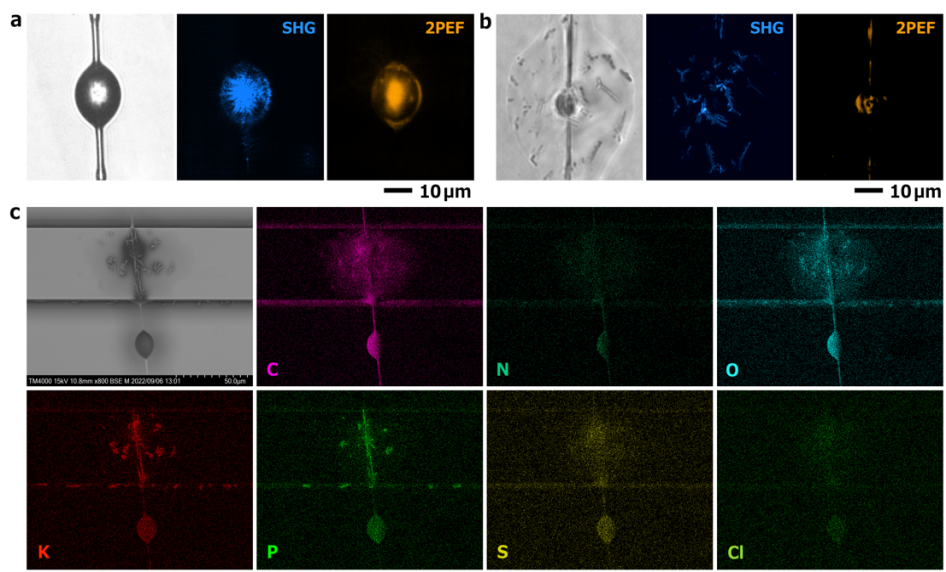

**Figure 4 | Second harmonic generation (SHG), 2-photon excitation fluorescence (2PEF) imaging and scanning electron microscope energy dispersive spectrometry (SEM-EDS) mapping of the aggregate glue droplet. a,** Suspended aggregate glue droplet microscopy images of white light illumination, SHG, and 2PEF. **b,** Attached aggregate glue droplet microscopy images of white light illumination, SHG, and 2PEF. **c,** SEM-EDS mapping images of the suspended and attached aggregate glue droplets. The color of SHG, 2PEF and SEM-EDS is a pseudo-color.

To investigate the behavior of aggregate glue droplets before and after adhesion, we performed selective observations of molecular structure and elemental mapping using second harmonic generation microscopy, 2-photon excitation fluorescence microscopy, and scanning electron microscope energy dispersive spectrometry. Figures 4a and 4b show microscopic images of second harmonic generation and 2-photon excitation fluorescence microscopy in suspended and attached aggregate glue droplets, respectively. Although the second harmonic generation, 2-photon excitation fluorescence in Fig. 4a, and the fluorescence signals in Fig. 1b are all scattered light, the distribution of the signals is different in the images due to different excitation mechanisms. Since the suspended aggregate glue droplet approximates a sphere, the second harmonic generation and 2-photon excitation fluorescence signals show a relatively strong image in the middle. Based on the shape of the suspended aggregate glue droplet, the substances producing second harmonic generation and 2-photon excitation fluorescence signals should be uniformly distributed. On the other hand, since the attached aggregate glue droplet approximates a thin film, the signal distributions of the 2-photon excitation fluorescence in Figs. 4b and 1e are consistent, *i.e.*, the fluorescent signal originates from the protein. Second-harmonic microscopy can selectively observe structures with broken inversion symmetry. The originally uniform second harmonic generation and 2-photon excitation fluorescence signals exhibited local intensity distributions after the suspended aggregate glue droplet was attached to the $CaF_2$ substrate. Previous reports have observed second harmonic generation and 2-photon excitation fluorescence signals in suspended polymerized gel droplets, but the origin of the nonlinear optical response is unclear [44]. The branching structure presented by the second harmonic generation signal in Fig. 4b suggests that the origin of the second harmonic generation is a crystal structure. Subsequently, the elements in the aggregate glue droplet were analyzed and mapped by scanning electron microscope energy dispersive spectrometry. Here, the aggregate glue droplet is attached to a grooved Si substrate. As shown in Fig. 4c, suspended and attached aggregate glue droplets are observed simultaneously using the step of the substrate and groove. In the element mapping images, either suspended or attached aggregate glue droplet, C, N, S, Cl elements did not show local distribution, but in the attached area, the intensity distribution of K, P, O elements showed a branching structure similar to second harmonic generation signals. This result suggests that the origin of second harmonic generation is a crystal containing K, P, O elements. Therefore, the origin of the second harmonic generation in the aggregate glue droplet can be determined to be a potassium dihydrogen phosphate crystal, which is noted for its non-linear optical properties [45]. In summary, the spread of the aggregate glue droplet after it is attached to the substrate rapidly increases its surface area, which breaks the original vapor-liquid equilibrium. Evaporation of water results in the deposition and crystallization of potassium dihydrogen phosphate. On the contrary, since the suspended aggregate glue droplets were already in the vapor-liquid equilibrium, when the vapor pressure of water did not satisfy the conditions for the deposition of potassium dihydrogen phosphate, no SHG signal was observed in the suspended aggregate glue droplets.

**Discussion**

Glycoproteins and low molecular mass compounds are uniformly distributed in the suspended aggregate glue droplet. Glycoproteins dissolve in low molecular mass compounds and exhibit fluidity. In this study, first, the phase separation of the originally uniformly distributed low molecular mass compounds and glycoproteins in the aggregate glue droplet after adhesion was confirmed morphologically by fluorescence microscopy, and the fluorescence of the aggregate glue droplet was determined to come from glycoproteins. Once the aggregate glue droplet is contacted by a prey or substrate, low molecular mass compounds rapidly separate from glycoproteins and diffuse along the prey or substrate. Glycoproteins harden upon loss of solvent consisting of low molecular mass compounds. Subsequently, the spatial distribution of various components of low molecular mass compounds in aggregate glue droplets after attachment was characterized and mapped using TOF-SIMS. As a result, the distribution of choline and dihydrogen phosphate was consistent. The mapping results of mid-infrared absorption confirm that choline dihydrogen phosphate in low molecular mass compounds acts as an ionic liquid to dissolve proteins in aggregate glue droplets. Therefore, we concluded that the constituents of aggregate glue droplets include an ionic liquid: hydrated choline dihydrogen phosphate, one of whose function is to dissolve proteins. By comparing the second harmonic generation microscopy images and scanning electron microscope-energy dispersive X-ray spectrometry elemental mapping results, the origin of SHG in aggregate glue droplets was determined to be the precipitated potassium dihydrogen phosphate crystal. Therefore, the selective observation of potassium dihydrogen phosphate by second harmonic generation microscopy can directly characterize the deposition or dissolution of dihydrogen phosphate, and indirectly characterize the concentration of ionic liquid of hydrated choline dihydrogen phosphate in aggregate glue droplets. As a result, the hydrated ionic liquid in low molecular mass compounds controls the balance of their concentrations by precipitating or dissolving potassium dihydrogen phosphate when suspended aggregate glue droplets are not in contact with any foreign bodies. After the aggregate glue droplets come into contact with the substrate, the evaporation of water leads to an increase in the concentration of ionic liquid, which in turn leads to the precipitation of potassium dihydrogen phosphate crystals. The retention of protein fluorescence activity before and after the attachment of aggregate glue droplets indicates that the ionic liquid can preserve its higher-order structures. The premise of the stable properties of proteins in the ionic liquid is that choline dihydrogen phosphate is in thermodynamic equilibrium, that is, the modification of glycoproteins in aggregate glue droplets depends on the concentration of ionic liquid.

The mechanism of spider orb web adhesion can be summarized as once aggregate glue droplets touch the object, their surface area expands rapidly, and the evaporation of water breaks the original concentration balance and makes $H_2PO_4^-$ in hydrated choline dihydrogen phosphate binds to $K^+$ and precipitates. An increase in the concentration of the ionic liquid of choline dihydrogen phosphate reduces its solubility for proteins, which in turn leads to the hardening of the glycoproteins of aggregate glue droplets. After protein denaturation, the physicochemical properties change, such as the decrease of solubility and the formation of precipitation. Because some of the hydrophobic groups originally inside the molecule are exposed due to the loosening of the structure, the asymmetry of the molecule increases, thus the viscosity increases and the diffusion coefficient decreases. Aggregate glue droplets effectively capture prey through this mechanism.

Ionic liquids have huge potential applications in pharmaceuticals, biopolymer dissolution processing, coating material for insulation samples for electron microscopes, matrix material for matrix-assisted laser desorption/ionization (MALDI), battery electrolyte, etc. As a natural ionic liquid, aggregate glue droplets from the spider orb webs may be an excellent Ionic liquid material model.

# Methods

## Sample collection

Sample collection was carried out during the night at July to September in the 5 years from 2018 to 2022, on the road along the Kawaguchi River (coordinates (WGS84): E139.31, N35.68), Hachioji, Tokyo, and around the Toyota Technological Institute campus (coordinates (WGS84): E136.98, N35.11), Nagoya. Samples were obtained from the orb web of the spider *Neoscona nautica* (see Supporting Information Fig. S7). After the sample was brought back to the laboratory, the capture spiral line silks were attached to a slide glass for visible light microscopic observation, and attached to a $CaF_2$ substrate for second harmonic microscopic, mid-infrared imaging and scanning electron microscope-energy dispersive X-ray spectrometry (SEM-EDS) measurement. For TOF-SIMS measurement, the capture spiral line silks were attached to a piece of aluminum with a diameter of 10 mm. Then the measurement was performed immediately. In all analyses, the sample is pristine, without any chemical treatment (e.g., washing, separation, etc.).

## Fluorescence microscopic observation

A microscope (Olympus IXplore Standard) was used for fluorescence microscopic observation. A white light source was used for observing a normal microscopic image. The fluorescence was excited by an ultraviolet light source. The spectrum of excitation ultraviolet light is shown in Supporting Information Fig. S8. The energy density of excitation ultraviolet light was about 42.3 mJ/mm². The fluorescence signal was observed with a long-wavelength pass filter with a cut-on wavelength at 420 nm for blocking the excitation light. The optical spectrum was measured by a fiber spectrometer (Ocean Optics HR2000+).

**FIB-TOF-SIMS**

A home-made FIB-TOF-SIMS instrument [22] was used for component analysis and elemental mapping of aggregate glue droplets. The TOF-SIMS analyses were operated at a temperature of 22˚C and a pressure of $3.2 \times 10^{-6}$ Pa (with sample). The ion source for the FIB is $Ga^+$, the acceleration voltage is 30 keV, and the beam angle is 45˚. TOF-SIMS analyses used a pulse mode. The beam current (DC) was set to 1.8 nA and the FIB pulse time width is set to 200 ns with 10 kHz. Delayed extraction is adopted. The mapping was performed by scanning the sample surface with the FIB. In elemental mapping, the field of view was 500 μm × 500 μm. The mass resolution is $m/\Delta m = 4000$ (at $m/z$ 28).

**MIR imaging**

A high speed full-field entire bandwidth mid-infrared hyperspectral imaging system [35] was used for the chemical imaging of aggregate glue droplets. Our past reports describe the principles and setup of mid-infrared hyperspectral imaging in detail [35], here we show the specific parameters in this study. The whole sample is excited with a full-field irradiation by the MIR pulses with a pulse width of 13.6 fs which is sub-half-cycle phase-stable pulses and a spectral bandwidth of 3–30 μm. After the sample, a chirped pulse with a duration of 1.8 ps and a center wavelength of 797 nm was introduced and is coaxialized with the transmitted MIR beam. A nonlinear crystal GaSe film was placed on the image plane, and the MIR pulses was converted to visible light. Since the chirped pulse is used for the upconversion, the MIR pulse with a tail of picosecond free induction decay (FID) signals is fully converted to visible, and the information of the MIR spectrum is maintained with the resolution of the inverse of the chirped pulse duration. The upconverted light is detected with a hyperspectral camera, and a hyperspectral image, namely the spectral image data cube, is obtained. The hyperspectral image data was obtained taking 8 s. The data cube consists of images with 1069 wavelength points, which corresponded to a wavenumber resolution of 3 cm$^{-1}$. Finally, the components of the sample are mapped by spectral angle mapper analysis and linear discriminant analysis.

**Second harmonic generation and 2-photon excitation fluorescence microscope**

The second harmonic and 2-photon excitation fluorescence microscopy observation by a home-made microscope. The optical setup is shown in Supporting Information Fig. S9. The light source was a Ti:sapphire laser system based on a regenerative amplifier and a single-pass amplifier (Spectra-Physics Spitfire Ace) with a repetition frequency of 5 kHz, a center wavelength of 800 nm, and a pulse width of 35 fs. The fundamental beam was focused by a lens ($f$ = 200 mm) and irradiated the sample from directly below. The diameter of the beam spot on the sample plane was 0.8 mm$^2$, and excitation light energy density of one pulse was about 0.1 mJ/mm$^2$. Since the irradiated area is much larger than the sample size, the excitation energy distribution in the sample is substantially uniform. The power of the excitation beam was controlled by a continuous variable neutral density filter before the sample. The scattered light from the sample passed through an infinity corrected objective lens (OLYMPUS, SLMPLN50x, NA=0.35) became a parallel ray, passed through a high reflector at 800 nm wavelength, rejecting 800 nm wavelength, and finally was selected by a bandpass filter (Semrock, FF01-395/11) with a center wavelength of 400 nm for the second harmonic light. Here, changing the final filter into a longpass filter (Semrock, FF01-430/LP-25, cut-on wavelength 437 nm), the 2-photon excitation fluorescence signal can be selected. Finally, the parallel rays pass through an imaging lens and the image was observed using a CMOS camera

(Hamamatsu, ORCA-Fusion BT, C15440-20UP, number of effective pixels: 2304 × 2304, pixel size: 6.5 µm × 6.5 µm). Each image taked an integration time of 500 ms. In the case of suspended aggregate glue droplets, the sample was fixed between two steel wires using its own viscosity. In the attached case, the aggregate glue droplets were attached to a $CaF_2$ substrate.

**SEM-SED**

A scanning electron microscope (HITACHI, TM4000PlusII) was used for energy dispersive X-ray spectrometry (Oxford Instruments, AZtecLiveOneXplore) measurement. The sample of aggregate glue droplets were attached to a $CaF_2$ substrate. The accelerating voltage for the scanning electron microscope image and the dispersive X-ray spectrometry analysis was 15 kV.

**Reagent list**

Choline dihydrogen phosphate, CAS RN: 83846-92-8, Kanto Chemical Co., Inc.

**Acknowledgements**

This work was supported by JSPS KAKENHI Grant Number JP21K14556. The mid-infrared hyperspectral chemical imaging system of this work was supported by Japan Science and Technology Agency (JST) Core Research for Evolutional Science and Technology (CREST), Grant Number JPMJCR17N5), Japan.

**Contributions**

Y.Z conceived the concept and supervised the work. Y.Z. prepared the sample and performed the experiment. Y.Z. and T.F. designed and built the mid-infrared hyperspectral chemical imaging system. T.S. designed and built the TOF-SIMS equipment. Y.Z. analyzed the data and wrote the manuscript. All the authors contributed to the discussion of the results and the manuscript.

**Corresponding author**

Correspondence to Yue Zhao.

ORCID: 0000-0002-8550-2020

e-mail: zhaoyue@toyota-ti.ac.jp

**Competing interests**

The authors declare no competing interests.

*Supplementary Information*

# A natural ionic liquid: low molecular mass compounds of aggregate glue droplets on spider orb webs

Yue Zhao [1,2,*], Takao Fuji [1], Masato Morita [3], Tetsuo Sakamoto [3]

[1] *Laser Science Laboratory, Toyota Technological Institute, 2–12–1 Hisakata, Tempaku-ku, Nagoya 468–8511, Japan.*
[2] *Collaborative Open Research Center, Kogakuin University, 2665-1 Nakano, Hachioji, Tokyo, 192-0015 Japan*
[3] *Department of Applied Physics, School of Advanced Engineering, Kogakuin University, 2665-1 Nakano, Hachioji, Tokyo, 192-0015 Japan*

* Correspondence to Yue Zhao.
   e-mail: zhaoyue@toyota-ti.ac.jp

# List of Contents

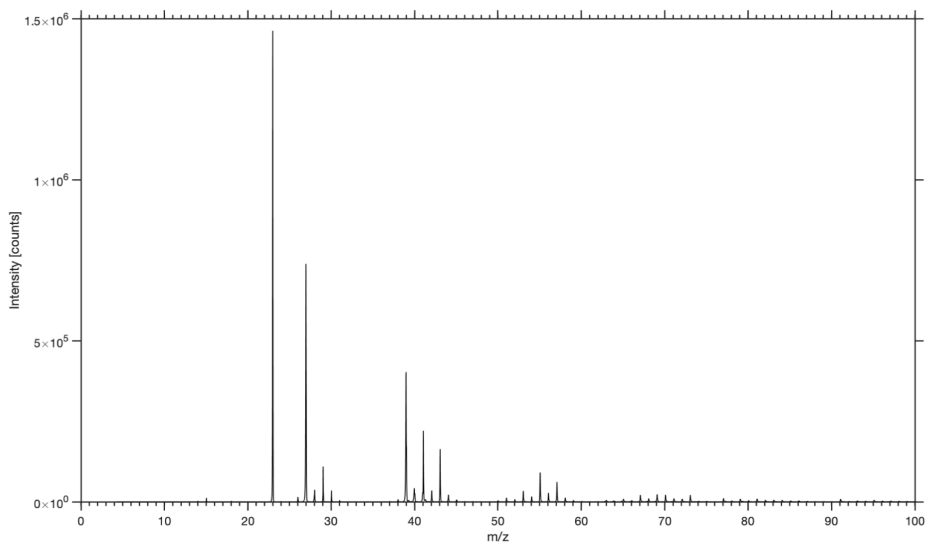
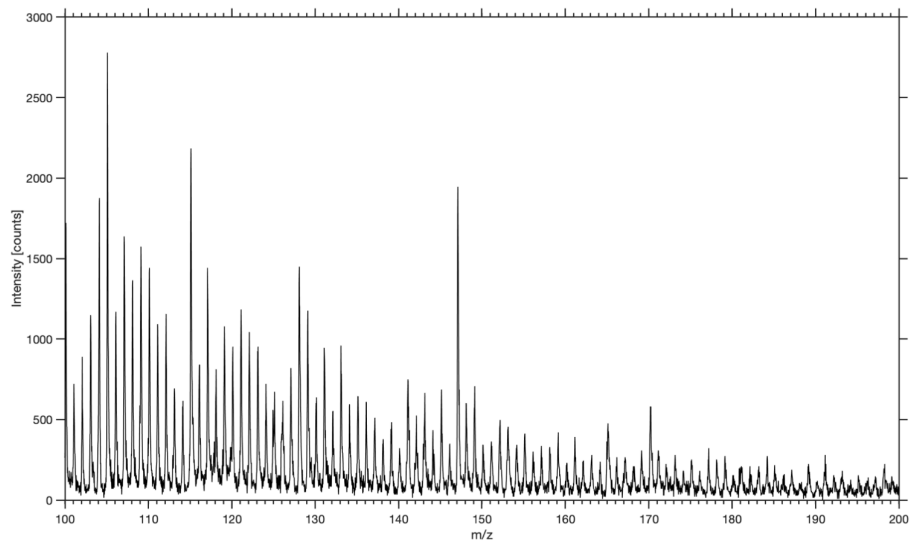
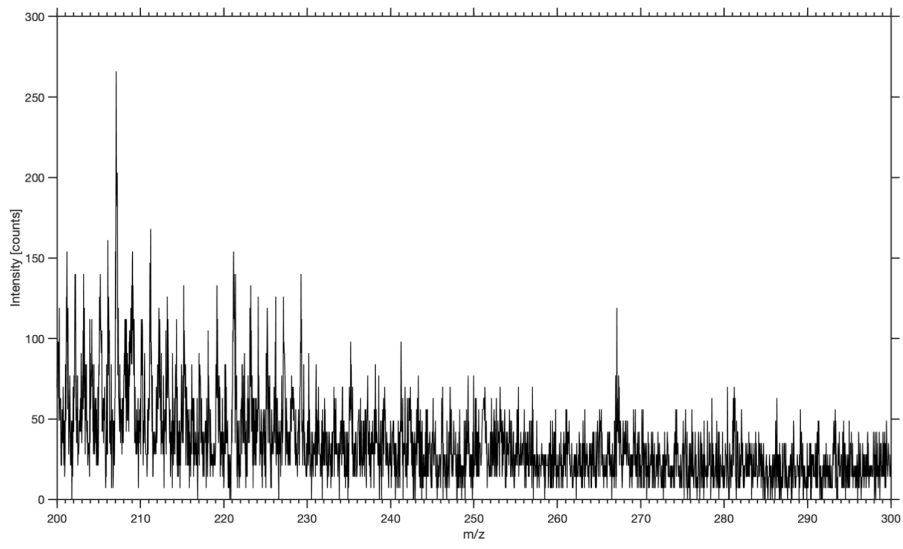

**Supplementary Figure S1** Mass spectrum (m/z 1 to 300) of positive ions of the aggregate glue droplets attached to the substrate.

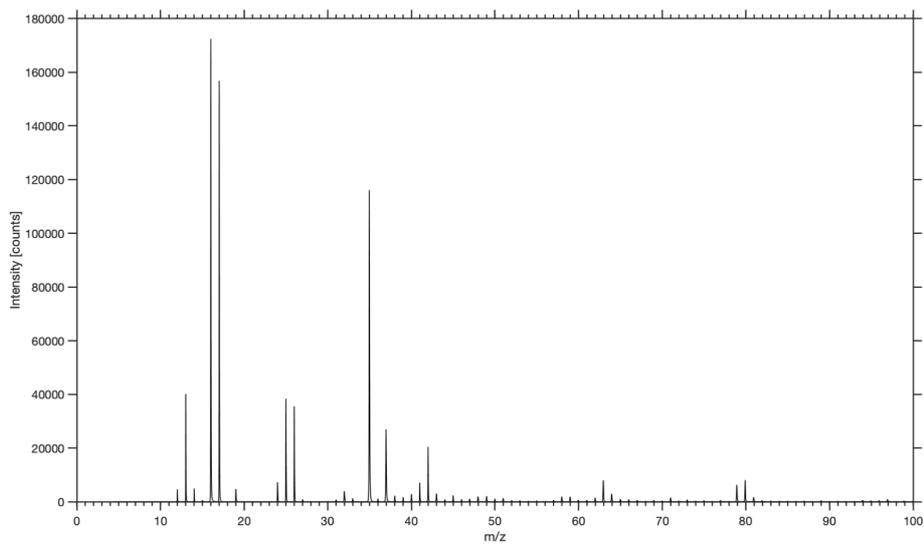
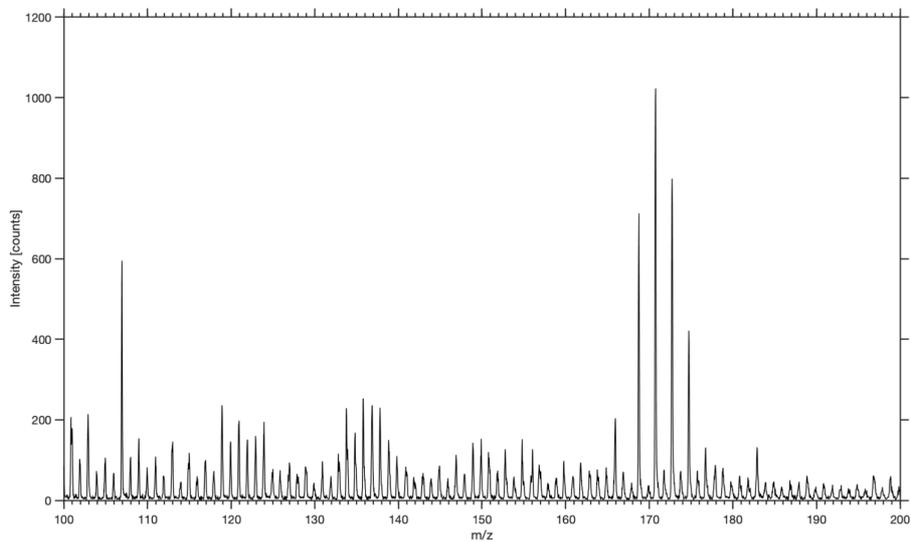
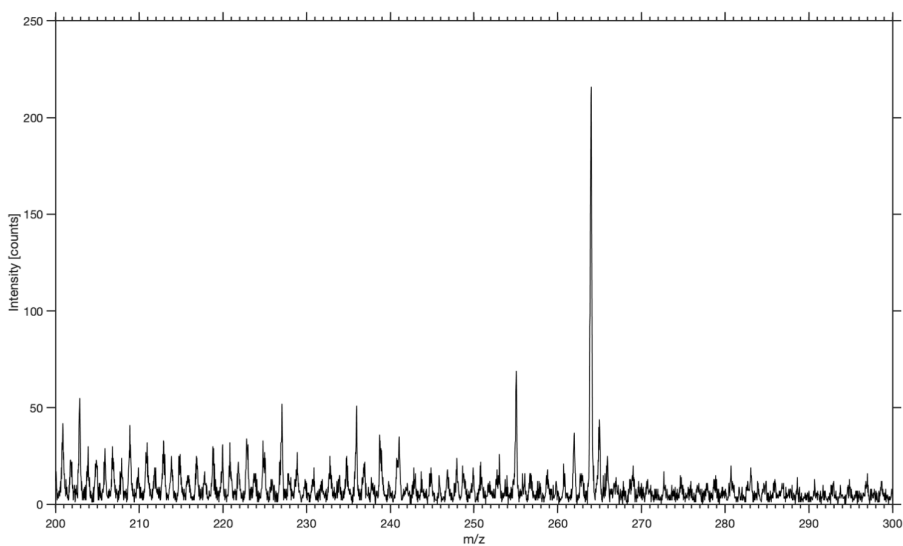

**Supplementary Figure S2.** Mass spectrum (m/z 1 to 300) of negative ions of the aggregate glue droplets attached to the substrate.

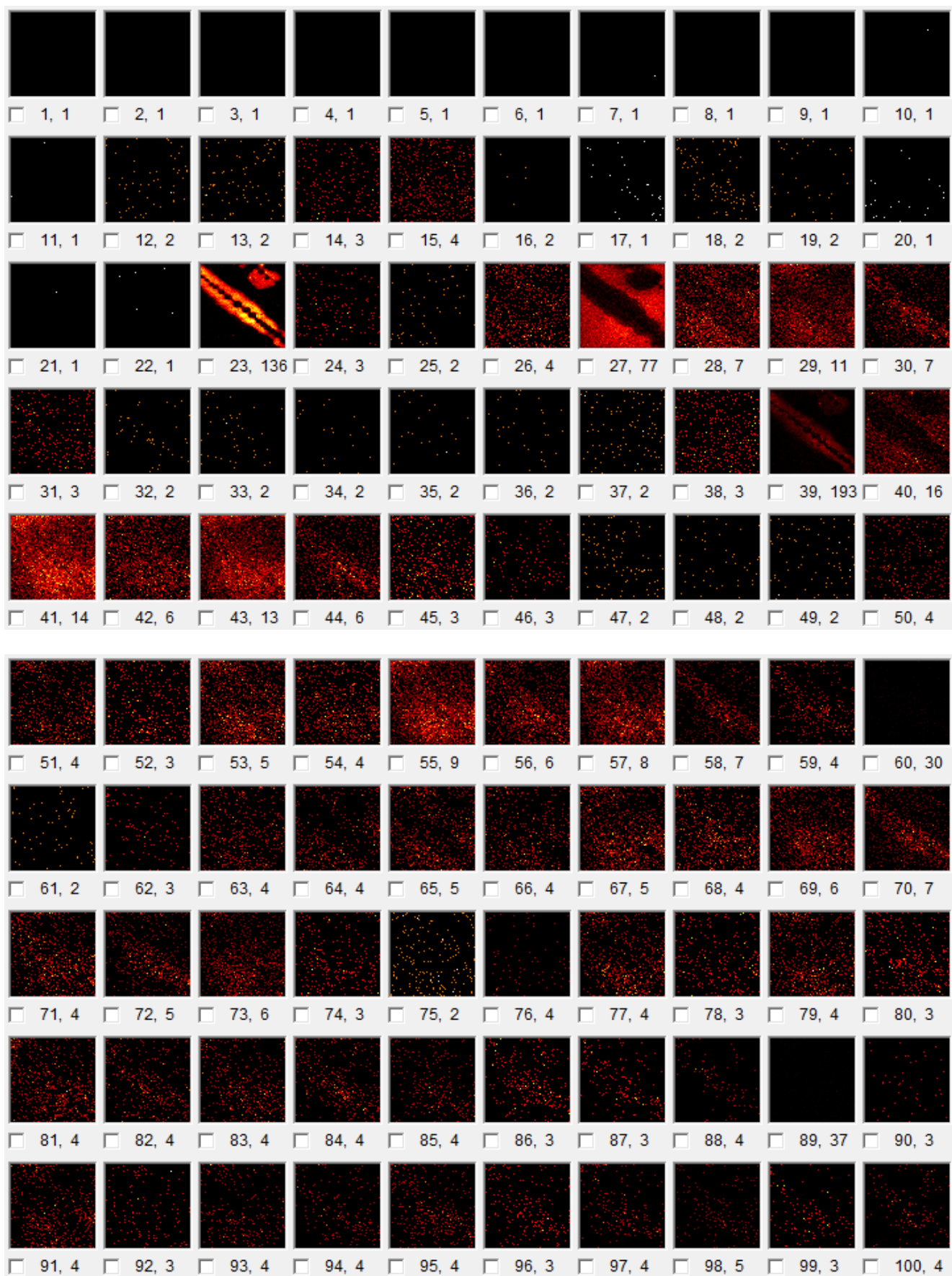

**Supplementary Figure S3.** Secondary positive ion maps (m/z 1 to 300) of the aggregate glue droplets attached to the substrate. The numbers under the maps represent *m/z*, max count.

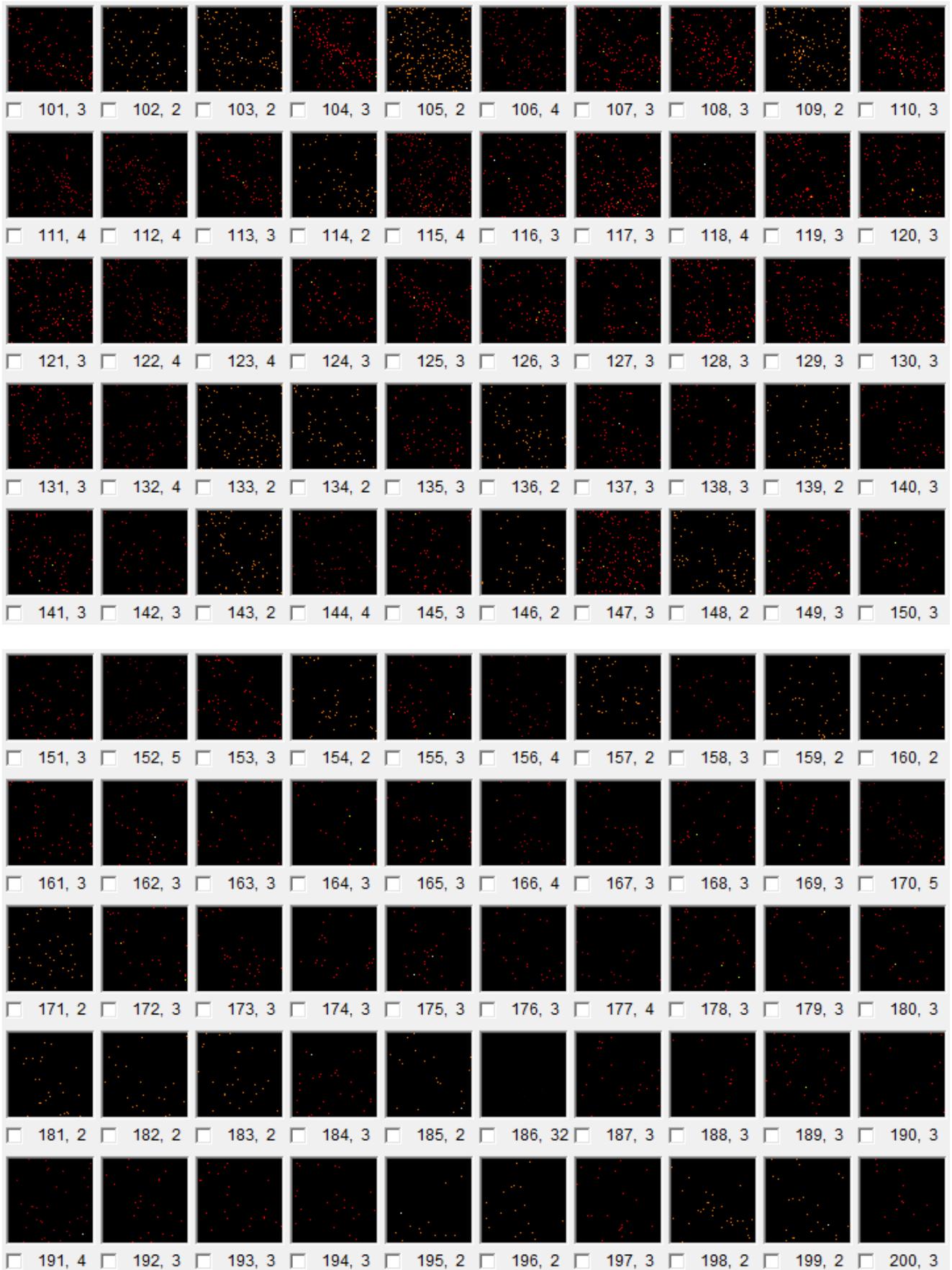

**Supplementary Figure S3.** (*Contd.*)

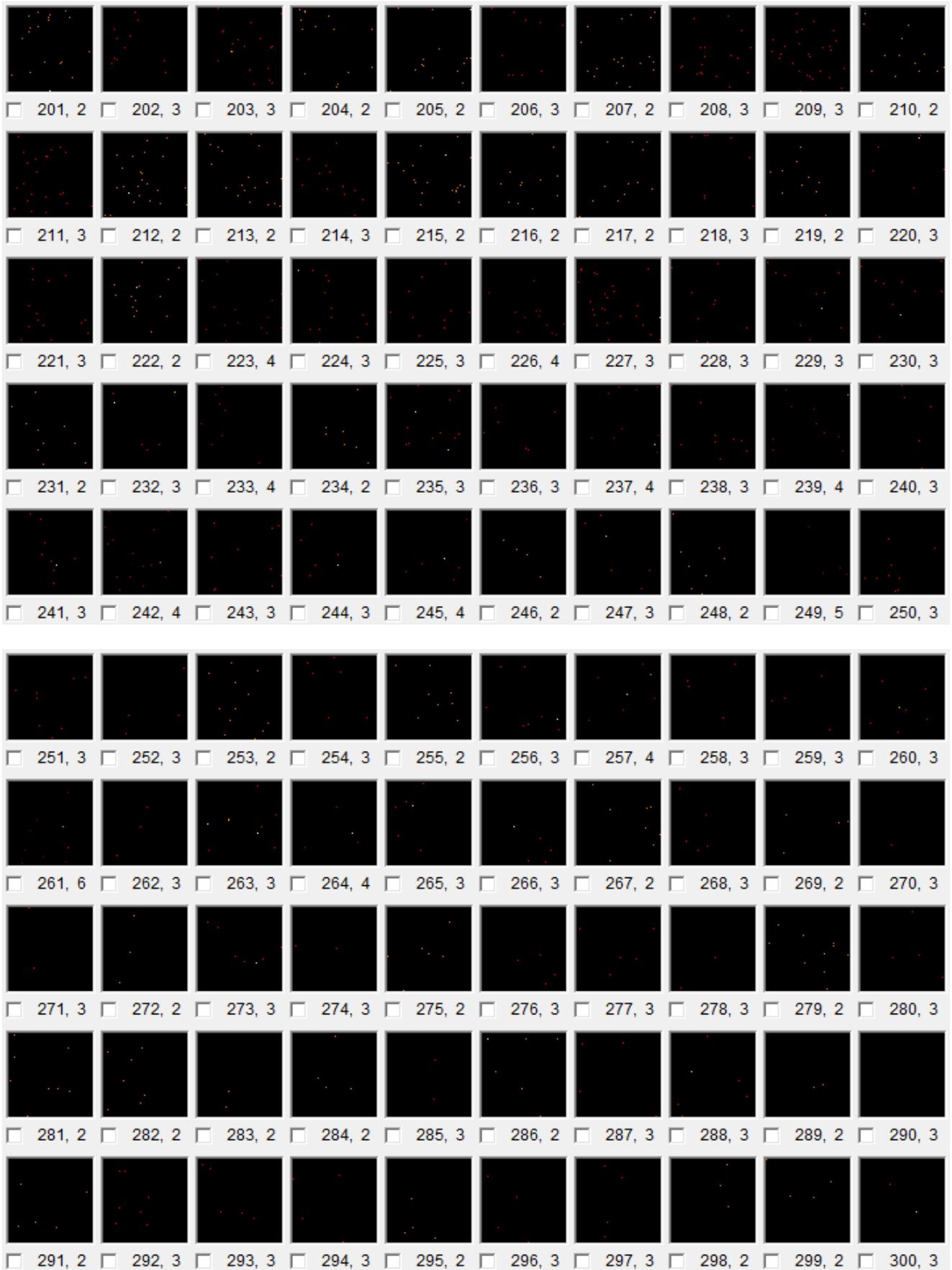

**Supplementary Figure S3.** (*Contd.*)

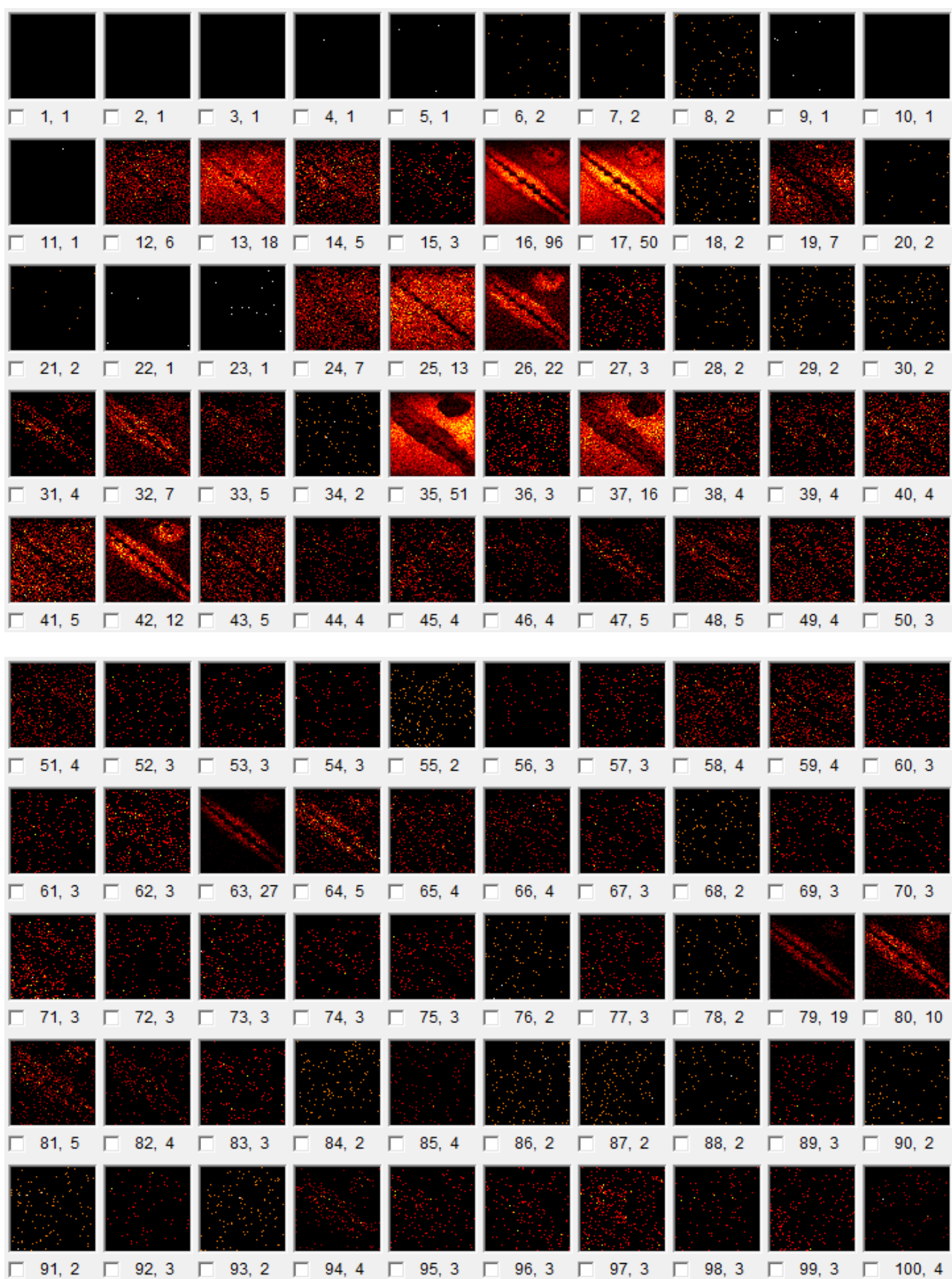

**Supplementary Figure S4.** Secondary negative ion maps (m/z 1 to 300) of the aggregate glue droplets attached to the substrate. The numbers under the maps represent *m/z*, max count.

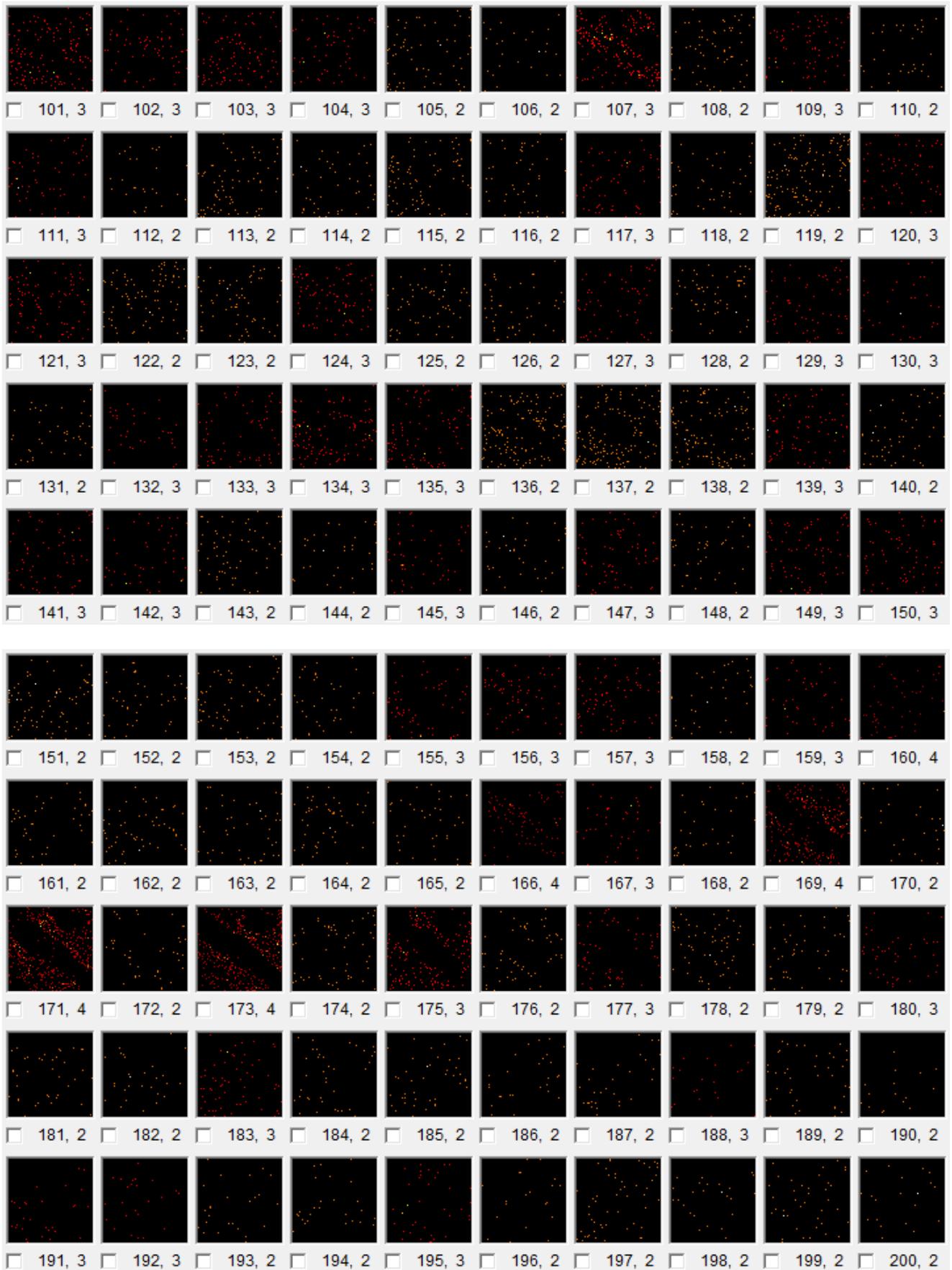

**Supplementary Figure S4.** (*Contd.*)

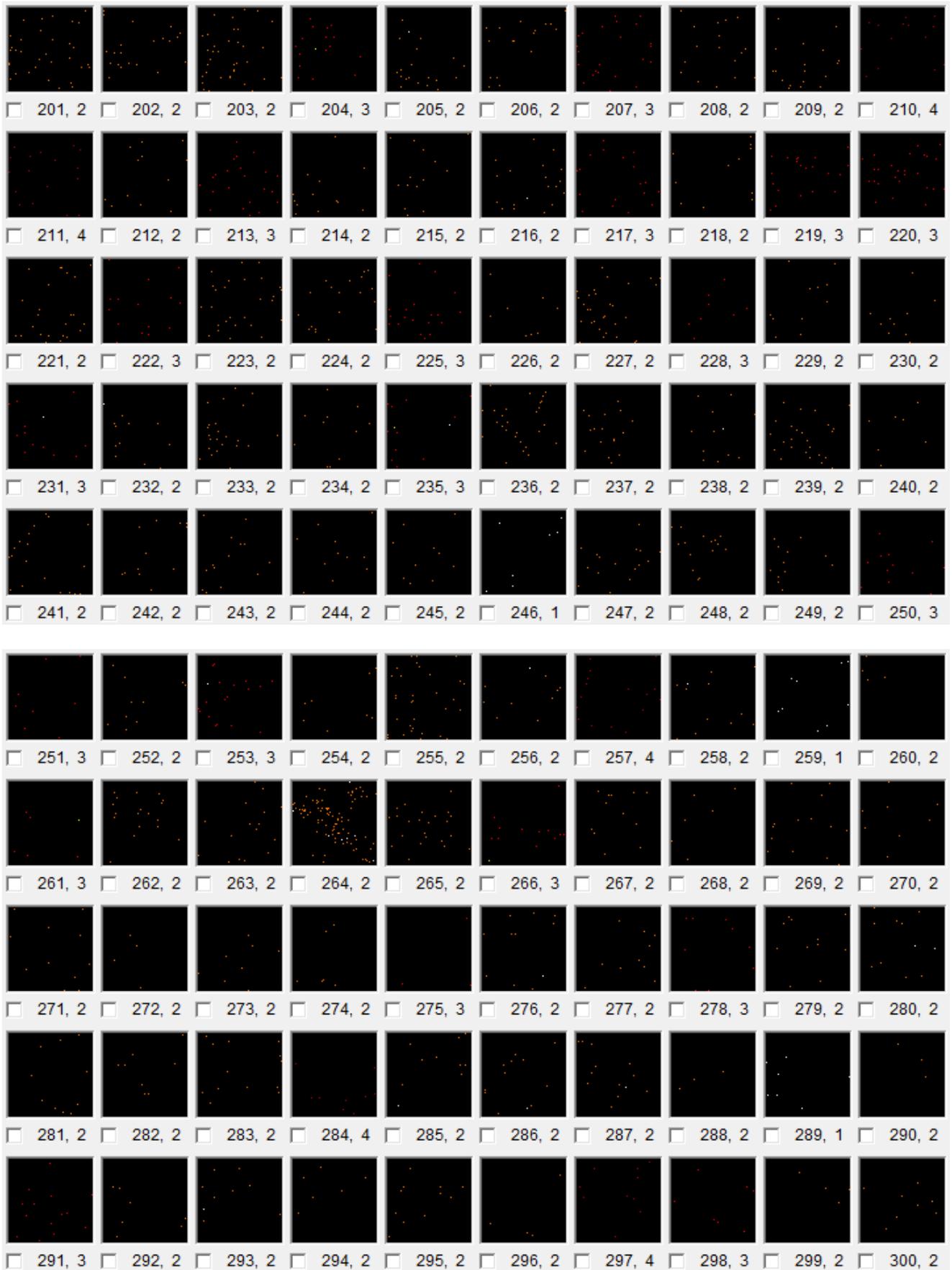

**Supplementary Figure S4.** (*Contd.*)

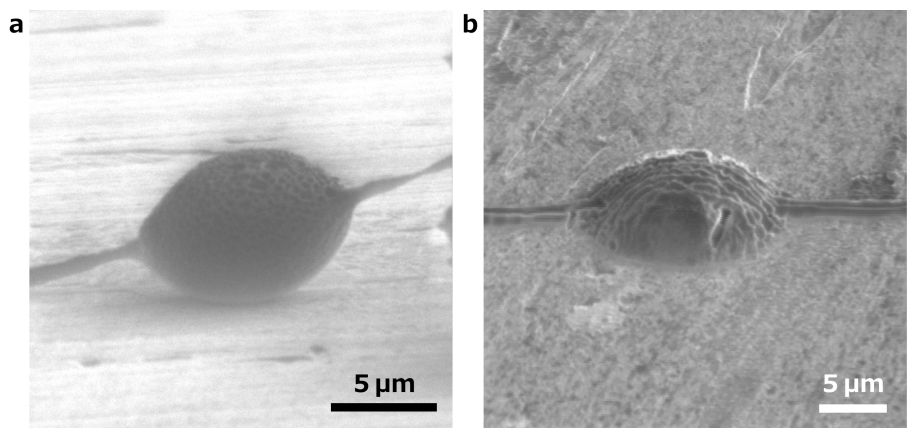

**Supplementary Figure S5.** Topographic nanopatterns of **a,** the suspended and **b,** the attached aggregate glue droplets by ion beam irradiation.

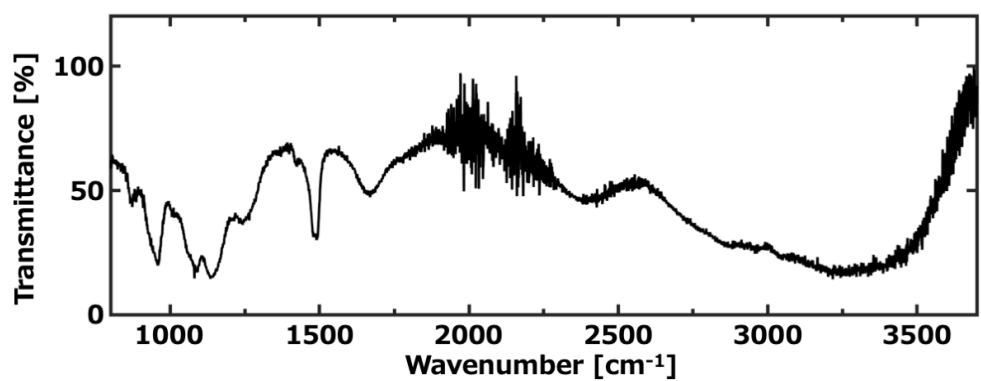

**Supplementary Figure S6.** Corrected infrared transmission spectrum of choline dihydrogen phosphate

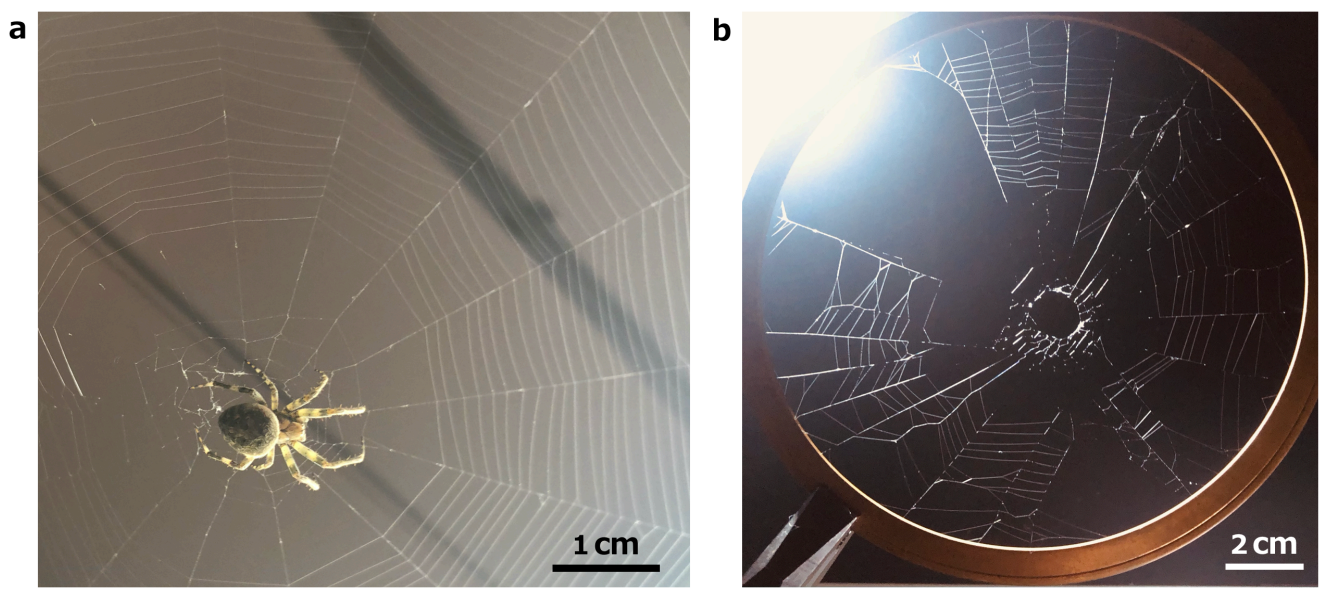

**Supplementary Figure S7. a,** A spider (*Neoscona nautica*). **b,** The sample of an orb web.

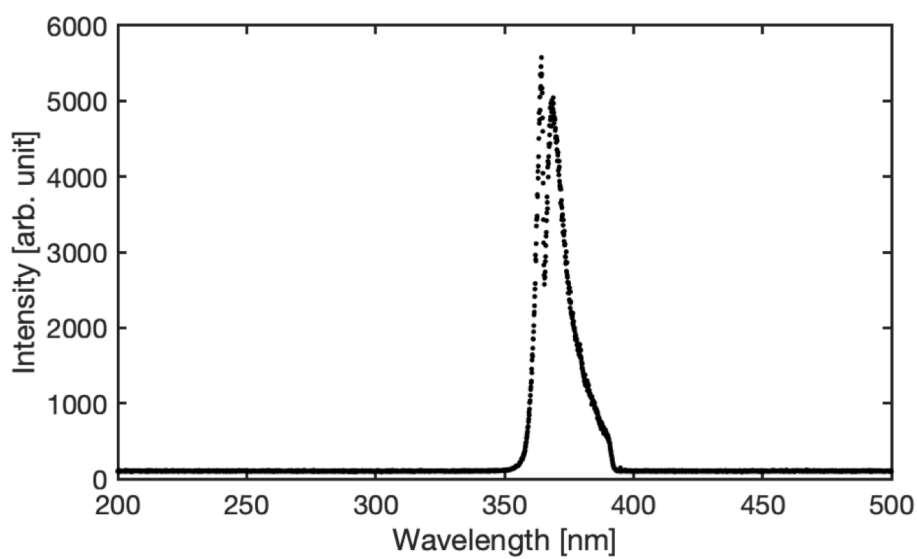

**Supplementary Figure S8.** The spectrum of excitation ultraviolet light.

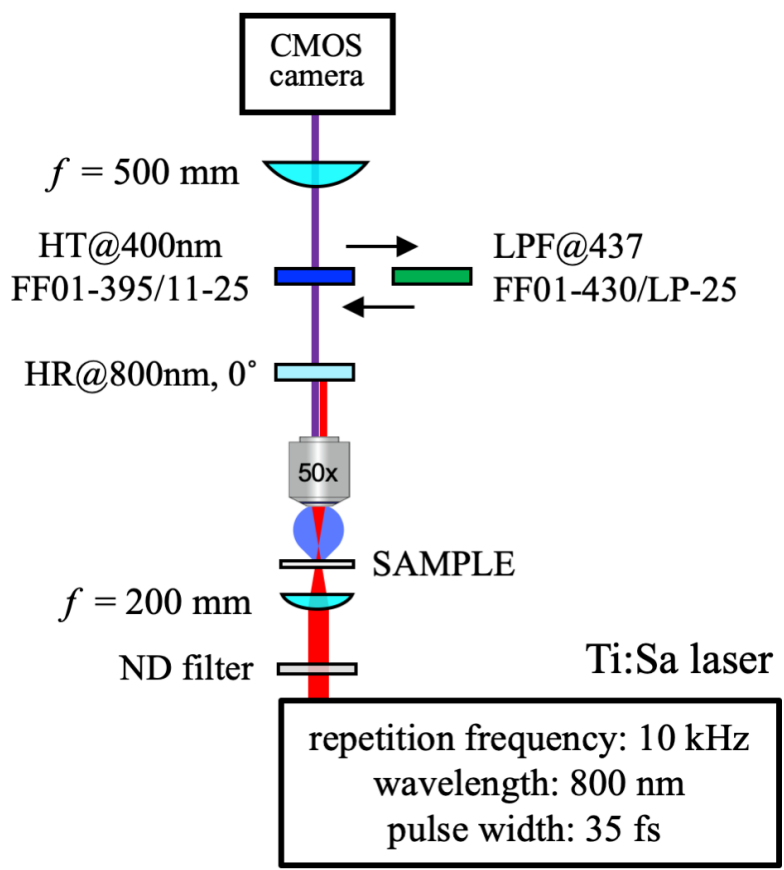

**Supplementary Figure S9.** Optical setup of the second harmonic generation imaging system.

11